\pdfoutput=1
\documentclass[aps,prl,reprint,superscriptaddress,amsmath,floatfix]{revtex4-1}
\usepackage[english]{babel}
\usepackage{amsmath,amssymb}
\usepackage{bm}
\usepackage{ifpdf}
\usepackage{graphicx}
\usepackage{float}
\usepackage{amsthm}
\usepackage{braket}
\usepackage{mathtools}
\usepackage[hypertexnames=false,bookmarks=false,colorlinks=true,linkcolor=blue,citecolor=blue]{hyperref}
\newcommand{\numberthis}{\addtocounter{equation}{1}\tag{\theequation}}
\usepackage{version}	
\begin{document}
\title{Suppression of superfluidity by dissipation \\
--- An application to failed superconductor }

\author{Kou Misaki,${}^{1}$ Naoto Nagaosa${}^{1,2}$}

\affiliation{
$^1$Department of Applied Physics, The University of Tokyo, 
Bunkyo, Tokyo 113-8656, Japan
\\
$^2$RIKEN Center for Emergent Matter Science (CEMS), Wako, Saitama 351-0198, Japan
\\
}
\date{\today}
\begin{abstract}
The ground states of bosons have been classified into superfluid, Mott insulator, and bose glass. 
Recent experiments in two-dimensional superconductors strongly suggest the existence 
of the fourth quantum state of Cooper pairs, i.e., bose metal or quantum metal, where the resistivity remains constant at lowest temperature. However, its theoretical understanding remains unsettled. In this paper, we show theoretically that the bosons in the dilute limit subject to dissipation can lose the superfluidity and remain metallic, utilizing the Feynman's picture of superfluidity in the first quantized formulation. 
This result is relevant to the quantum vortices under an external magnetic field in two-dimensional superconductors with the finite resistivity of the normal core as the source of dissipation.
\end{abstract}
\maketitle
{\it Introduction.--- }
Recent experiments show the possible metallic state down to 
the lowest temperature with variable resistivity
in two-dimensional superconductors,  in sharp contrast
to the conventional picture that the metallic state appears only at the
quantum critical point between the insulator and superconductor 
\cite{kapitulnik_colloquium:_2019}. Since we are interested in the
temperature region much lower than the superconducting gap,
the Cooper pairs can be regarded as charge $2e$ bosons, 
and hence the problem is regarded as that of the bosons.
Vortices, either due to quantum fluctuation or by external magnetic field,
play crucial role in this problem, and the insulating state/ superfluidity state
of the vortices corresponds to superconducting/insulating state of 
the Cooper pairs in the duality picture.
 
There are many papers on the dissipative XY model 
\cite{dalidovich_fluctuation_2000,zhu_local_2015,zhu_quantum_2016,hou_phase_2016}, describing the dynamics of the resistively shunted Josephson junction array \cite{chakravarty_onset_1986,chakravarty_quantum_1988}. 
However, the XY model is an effective model of bosons only at integer fillings \cite{fisher_quantum_1988,fisher_boson_1989}. Away from integer fillings, e.g., in the dilute limit, the action contains the first order time derivative term \cite{sachdev_quantum_2011}, which is complex and invalidates the Monte Carlo study in the phase representation.
This difference is important in the context of the positive magnetoresistance of failed superconductor, since in the dilute limit the number of vortices change continuously as we increase the magnetic field.
Also, the effect of dissipation on dilute boson system has been studied in Ref. \onlinecite{cai_identifying_2014}, but their study is only for one dimensional system. The analytical argument we will discuss here is different from their argument relying on bosonization which is valid only in one dimensional system.

The bosonic system at zero temperature is known to be a perfect 
superfluid, i.e., $\rho_s=\rho$, where $\rho_s$ is the superfluid density and $\rho$ is the total particle density, if the system does not break the Galilean invariance \cite{greiter_hydrodynamic_1989,son_general_2006} (For a similar discussion in the case of the superconductivity, see \cite{aitchison_effective_1995,aitchison_finite-temperature_2000}.). The point of the argument is that, if we write down the effective action for the phase variable $\phi$ \cite{popov_functional_1987}, the Galilean invariance enforces the action to be the functional of only $\partial_t\phi-(\nabla\phi)^2/2m$. Since the coefficient of $\partial_t \phi$ in the effective action is the total particle number density, it enforces the coefficient of $(\nabla\phi)^2/2$ term to be the total density also, i.e., $\rho_s=n/m=\rho$. 
At finite temperature, since the imaginary time action at finite temperature is not Galilean invariant, the above argument does not apply.
Therefore, at finite temperature $\rho_s \neq \rho$ \cite{popov_functional_1987,aitchison_finite-temperature_2000}, in accordance to the Landau's famous expression of $\rho_s$ in terms of the thermal distribution of the quasiparticle \cite{stoof_ultracold_2009}. 

The Galilean invariance at $T=0$ is explicitly broken if we introduce the lattice potential or the disorder, leading to the Mott insulator \cite{imada_metal-insulator_1998} or the Bose glass \cite{fisher_boson_1989,huang_hard-sphere_1992,lopatin_thermodynamics_2002}.
Another possible source of the loss of the Galilean invariance is the nonlocal interaction along the time-direction which arises after we integrate out the gapless degrees of freedom. This depletion of superfluid component due to retarded interaction has been studied in Ref. \onlinecite{feigelman_two-dimensional_1993}, where the gapless degrees of freedom is the gauge field which mediates the interaction between vortices. 

In this paper, we will discuss the effect of the gapless degrees of freedom, i.e., the effect of the dissipation, on the bosonic many body system, in particular, the system with superfluidity. The effect of the dissipation is qualitatively different from the single particle case, since, because of the bose statistics, the world lines of bosonic particles can exchange their positions at $\tau=0$ and $\beta$, and, in the presence of dissipation, the action for this off-diagonal configuration of the world line is very different from the diagonal configuration \cite{weiss_quantum_2012}. Since the superfluidity is accompanied by the condensation of the exchange event \cite{feynman_atomic_1953,feynman_atomic_1954}, we expect the superfluidity is drastically suppressed. 
Below, we will give both analytical and numerical arguments to support this expectation. Then, we discuss another model where dissipation reduces the superfluid density. The fact that these two very different models show the reduction of the superfluidity indicates that the dissipation suppresses the superfluidity in general.

We also note that our model is relevant to the system with vortices in superconductor, since the vortex is known to behave as a boson, and the normal core acts as a source of dissipation if the energy level broadening of the bound states are larger than the energy level spacing \cite{kopnin_theory_2009}. We will discuss the consequence of our result in the context of vortices later.

 { \it Model.---}
The phenomenological action for the system of many bosons in the presence of the dissipation is,
\begin{align*}
    S&=\int_0^{\beta} d\tau \left(\sum_i \frac{m}{2}\dot{\vec{r}}+\sum_{i>j}V_{i,j}\right)\\
    &+\frac{\eta}{4\pi}
    \sum_i \int_0^{\beta}d\tau\int_0^{\beta}d\tau' \frac{\pi^2}{\beta^2}
    \left(\frac{\vec{r}_i(\tau)-\vec{r}_i(\tau')}{\sin\frac{\pi}{\beta}(\tau-\tau')}\right)^2, \numberthis \label{dissaction}
\end{align*}
where $i$ is the labeling of the bosons,
$V_{i,j}$ is the repulsive interaction between bosons, $m$ is the mass of the bosons, $\beta$ is the inverse temperature, and
the last term represents the effect of the Ohmic heat bath \cite{caldeira_quantum_1983,nagaosa_quantum_1999}.
We neglected the effective interaction between the bosons induced by the coupling to the heat bath
\cite{duarte_effective_2006}.

{\it Extended Feynman's argument.---}
Here, we argue the effect of the dissipative term from the perspective of Feynman's picture of superfluidity \cite{feynman_$ensuremathlambda$-transition_1953,feynman_atomic_1953}. In the first quantized form, superfluidity
is characterized by the presence of the macroscopically large exchange processes; it appears in the form of the large fluctuation of winding number \cite{pollock_path-integral_1987,ceperley_path_1995}.
In the absence of
the dissipation, if we assume that the effect of the repulsive interaction is simply renormalizing the mass of the bosons, the action for the macroscopic exchange process can be obtained from the single particle off-diagonal density matrix of the free particle, which is given by $y(|\bm{r}-\bm{r}'|) \propto \exp[-m(\bm{r} - \bm{r}')^2/(2\beta \hbar^2)]$, so the action is proportional to $\beta^{-1}$. 
Therefore, as $\beta\to\infty$, the entropy of the macroscopic exchange processes, which is constant as a function of temperature, overcomes the action for the exchange process, so the bosonic system shows superfluidity at finite temperature.
More concretely, following Feynman, we approximate the partition function of the system by the one of the problem of drawing polygons on a lattice and write it as $Z=\sum_L y(d)^L g(L)$, where $L$ is the number of links between vertices of the lattice, $d$ is the lattice constant, and $g(L)$ is the total number of the polygons with $L$ links.
Here we again note that $y$ can be approximated by the {\it off-diagonal} single particle density matrix, rather than the {\it diagonal} one as is used for the criterion of the superfluidity in a previous literature \cite{apenko_critical_1999}, although the Lindemann type criterion may be a good {\it necessary} condition for the superfluidity. In other word, what determines the action for the exchange is $\braket{p^2}$, the second moment of the momentum, rather than $\braket{r^2}$, the second moment of the position, since the off-diagonal density matrix represents the information of the momentum distribution through the Wigner transform as $y(|\bm{r}-\bm{r}'|)\propto \exp[-(\bm{r}-\bm{r}')^2\braket{p^2}/(2\hbar^2)]$. Here $y$ is gaussian since the Caldeira-Leggett action is quadratic, and we assume that this form remains valid even in the presence of the interaction between particles.
$\braket{p^2}^{-1}$ and $\braket{r^2}$ show drastically different behavior in the presence of the dissipation: The former remains constant down to $\beta\to \infty$, while the latter diverges as $\log\beta$ \cite{weiss_quantum_2012}. The reason for finiteness of $\braket{p^2}^{-1}$ was clearly explained by Caldeira and Leggett \cite{caldeira_quantum_1983}. To see this, we transform the last term in Eq. (\ref{dissaction}) as
\begin{align*}
    \frac{\eta}{4\pi} \int_{-\infty}^{\infty}d\tau' \int_0^{\beta}d\tau \left(\frac{\vec{r}_i(\tau)-\vec{r}_i(\tau')}{\tau-\tau'}\right)^2, \numberthis
\end{align*}
where the finite temperature kernel is replaced by the zero temperature kernel, but now we need to consider the interaction of the boson at $0 \leq \tau \leq \beta$ with the infinite family of ``image lines'', periodically extended from $0\leq \tau' \leq \beta$ to $-\infty \leq \tau' \leq \infty$. When we consider the process where $\vec{r}_i(0)\neq \vec{r}_i(\beta)$, i.e., the off-diagonal component of the single particle density matrix, 
the integral diverges because of the discontinuity at $\tau' = 0$ and $\tau'=\beta$; this divergence 
is regularized by the ultraviolet cutoff of the heat bath, but this contribution to the off-diagonal density matrix coming from the discontinuity remains finite even if we take the limit $\beta\to\infty$ \footnote{We note that there is a similar problem in the treatment of the effective mass of polaron \cite{feynman_slow_1955,khandekar_polaron_1988}.}.

\begin{figure}
    \centering
    \includegraphics[width=\columnwidth]{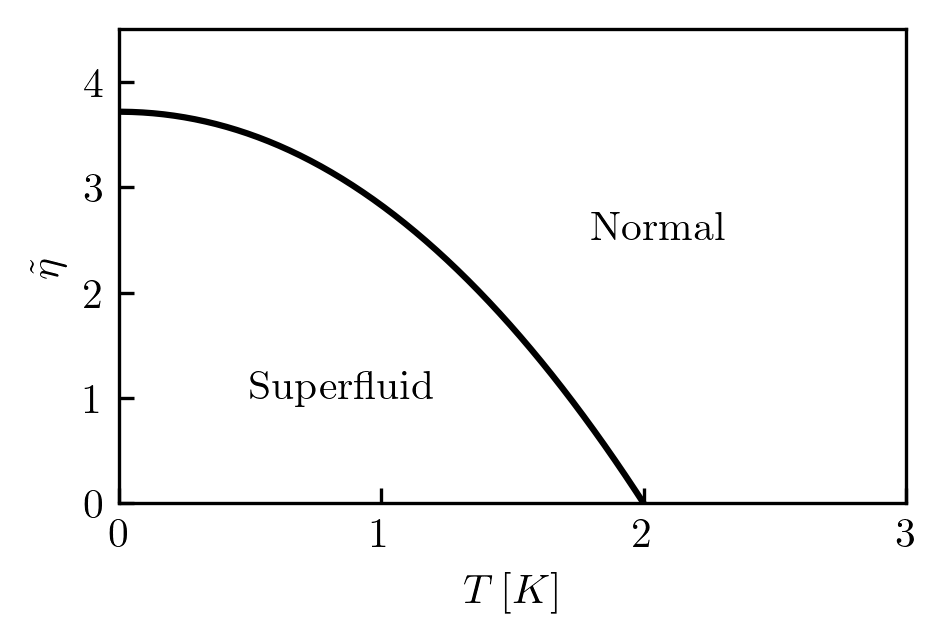}
    \caption{Schematic phase diagram obtained from Feynman's argument combined with the expression for the off-diagonal density matrix in the presence of the Ohmic dissipation, Eq. (\ref{exprforpsquare}). $\tilde{\eta}=\eta d^2$, where $d$ is the interparticle distance. The mass and the interparticle distance are the values for the Helium at saturated vapor pressure, and the cutoff for the bath, $\omega_D$, is set to be $10\,[{\rm K}]$. The phase boundary is calculated from the condition $\braket{p^2}(T,\tilde{\eta})=\braket{p^2}(T=2\,{\rm K},\tilde{\eta}=0)$, i.e., we assumed that the transition temperature for the dissipationless system is $T=2\,[{\rm K}]$.}
    \label{phasediagram}
\end{figure}

If we assume that the effect of the interaction can be renormalized to the effective mass of the particle, 
from the well known result of the quantum Brownian motion \cite{weiss_quantum_2012}, 
\begin{align*}
    \braket{p^2}&=\frac{M}{\beta}+ 2M \frac{\mu_1 \mu_2}{\mu_1-\mu_2} \left[\psi\left(1+\mu_1\beta\right)-\psi\left(1+\mu_2\beta\right)\right], \numberthis \label{exprforpsquare}
\end{align*}
where $M$ is the effective mass of bosons, $\psi(x)$ is the digamma function, $\mu_{1/2}=\hbar(\omega_D \pm \sqrt{\omega_D^2-4\gamma\omega_D})/(4\pi)$, $\gamma = \eta/M$, and $\omega_D$ is the cutoff of the spectrum of the bath.
Then, since $\braket{p^2}$ decreases as we lower the temperature and saturates at finite value, we expect that the transition temperature, which is the temperature where the entropy of macroscopic exchange $g(L)$ and the action for the exchange $y^L$ compete, monotonically decreases and reaches zero as we increases the coupling $\eta$. This behavior is schematically shown in Fig. \ref{phasediagram}. The critical $\eta$ at $T=0$ can be estimated from  $d^2\braket{p^2}_{T=0}/\hbar^2=2Md^2\mu_1 \mu_2[\ln{(\mu_1/\mu_2)}]/[\hbar^2(\mu_1-\mu_2)] \sim 1 $. If we further assume $\omega_D \gg \gamma$, the above condition simplifies to $\tilde{\eta} [\ln(\omega_D/\gamma)]/(\hbar \pi) \sim 1 $, where $\tilde{\eta}=d^2\eta$.

Below, we will show a strong support for this physical argument by the numerical Monte Carlo calculation of the superfluid density. This calculation confirms that the interaction between particles does not drastically affect the picture of superfluidity by Feynman even in the presence of the dissipation.

 {\it Result of the numerical calculation.--- }
We calculated the superfluid density for the boson system characterized by the action (\ref{dissaction}) 
with the worm algorithm in continuous space \cite{boninsegni_worm_2006-1,boninsegni_worm_2006}
using the winding number formula \cite{pollock_path-integral_1987,ceperley_path_1995}. We implemented the canonical version \cite{mezzacapo_superfluidity_2006,mezzacapo_structure_2007} where the Monte Carlo moves do not change the number of particles and employed
the Aziz potential \cite{aziz_accurate_1979} for the interaction. The convergence was checked by the binning analysis \cite{ambegaokar_estimating_2010}. Following Ref. \onlinecite{boninsegni_worm_2006}, we employed the Chin approximation 
\cite{chin_symplectic_1997} for the interaction term. The dissipative term was discritized as \cite{werner_quantum_2005,werner_phase_2005},
\begin{align*}
    &\frac{\eta}{2\pi}\sum_i\sum_{k>k'}\frac{\pi^2}{N_{\tau}^2}\frac{(\vec{r}_i(k)-\vec{r}_i(k'))^2}
    {\sin^2(\frac{\pi}{N_{\tau}}(k-k'))}\\
    \eqqcolon& \sum_i\sum_{k>k'}K(k-k')(\vec{r}_i(k)-\vec{r}_i(k'))^2,\numberthis
\end{align*}
where $N_{\tau}$ is the number of the Trotter step, $k$, $k'$ is the labeling of time slice. 
To avoid the divergence associated with the discontinuity at $k =0$ and $N_{\tau}$, we introduce the UV cutoff for $K$ as $K(k-k')=K((1-\tau_c)N_{\tau})$ for
$(1-\tau_c)N_{\tau} \leq k-k' \leq N_{\tau} - 1$; this form of cutoff is naturally 
realized if we introduce the ultraviolet cutoff for the spectrum of the heat bath.
\begin{figure}
    \centering
    \includegraphics[width=\columnwidth]{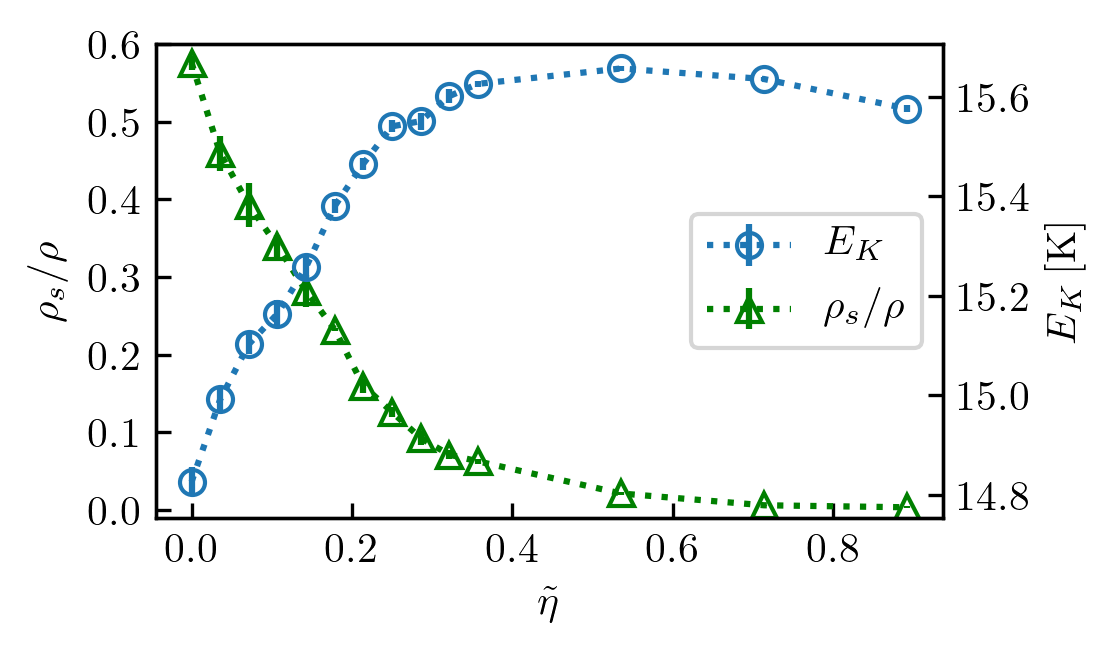}
    \caption{The superfluid fraction $\rho_s/\rho$ and the kinetic energy $E_K$ as a function of $\tilde{\eta}=\eta d^2$, where $d = 3.570\, {\rm \AA}$ is the interparticle distance. The blue circle represents the kinetic energy, while the green triangle represents the superfluid fraction.}
    \label{sfdenscalc}
\end{figure}
\begin{figure}
    \centering
    \includegraphics[width=\columnwidth]{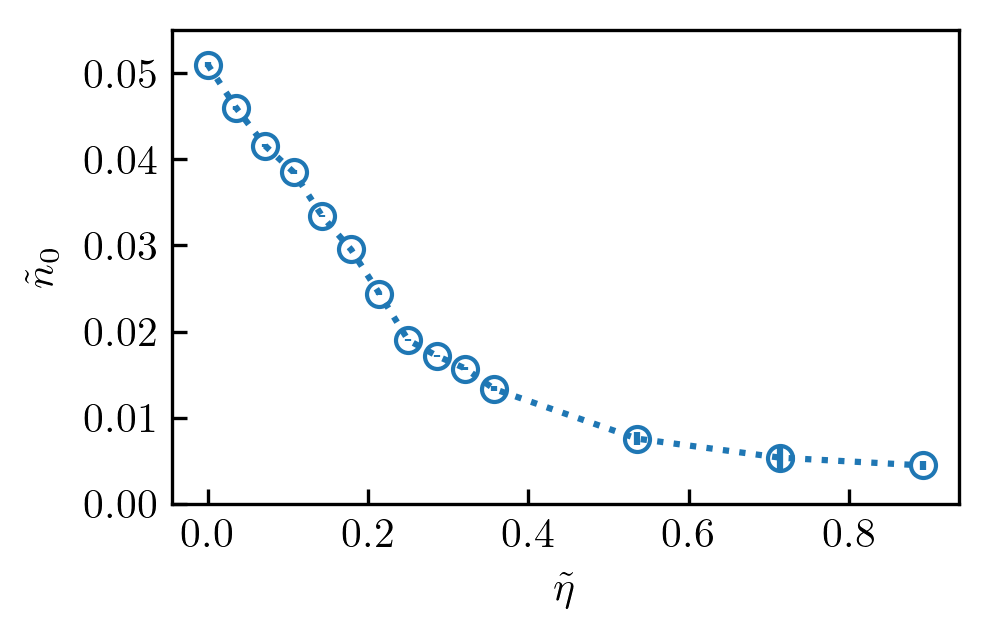}
    \caption{The condensate fraction at zero momentum, $\tilde{n}_0$, estimated from the off-diagonal density matrix.}
    \label{n0fig}
\end{figure}

We calculated the superfluid fraction for three dimensional system with the number of particles $N = 64$ at saturated vapor pressure, for $T=2\,{\rm K}$. The imaginary time step is $5\times 10^{-3}\,{\rm K^{-1}}$, and the cutoff of the bath is set to be $\tau_c=0.2$. The result of the calculation is shown in Fig. \ref{sfdenscalc} (green triangle). We can clearly see that $\rho_s$ monotonically decreases as a function of $\eta$. We also calculated the kinetic energy (blue circle), which characterizes how strong the bosons fluctuate in the imaginary time. We can see the increase of the kinetic energy as a function of $\eta$, which comes both from the suppression of fluctuation of each boson and the suppression of the exchange event, which lowers the kinetic energy \cite{ceperley_path_1995}.

Another important quantity is the off-diagonal density matrix, which can be easily calculated in the worm algorithm \footnote{See supplemental material for the details of the off diagonal density matrix and the numerical calculation in two dimension.}. We estimated the condensate fraction $\tilde{n}_0$ by fitting $n(r)$ with the function $\tilde{n}_0+(1-\tilde{n}_0)f(r)$, where $f(x)=\exp(-\bar{\alpha}_2 x^2/2!+\bar{\alpha}_4 x^4/4! - \bar{\alpha}_6 x^6/6!)$. 
This form of the fitting function is motivated by the one used in the absence of dissipation \cite{glyde_excitations_2017}. Here we ignored the contribution from the coupling term, since the form of the coupling term seems to be inapplicable in the presence of the dissipation. The ignorance of this term leads to an overestimation of $n_0$, but we believe that the qualitative trend as a function of $\eta$ can be captured by this simple fitting. The estimation of $\tilde{n}_0$ is shown in Fig. \ref{n0fig}. We can see the monotonic decrease of $\tilde{n}_0$ as a function of $\eta$.

{\it Second Model.--- }
Here, we discuss the effect of dissipation on the superfluidity in the following field theoretical model:
\begin{align*}
 S &=\sum_{\omega_n,\bm{k}}\left(-i\omega_n+\frac{\bm{k}^2}{2m}-\mu\right)\bar{\psi}_{n,\bm{k}}\psi_{n,\bm{k}}+\frac{g}{2}\int d\tau
d\bm{r}\bar{\psi} \bar{\psi} \psi \psi \\
&+\alpha\sum_{\omega_n,\bm{k}}|\omega_n|\rho_{\bm{k}}^n 
\rho_{-\bm{k}}^{-n},\quad (\rho_{\bm{k}}^n = \sum_{\omega_m,\bm{q}}\bar{\psi}_{n+m,\bm{k}+\bm{q}}\psi_{m,\bm{q}}),
\end{align*}
where $\psi,\bar{\psi}$ are the bosonic annihilation and creation operator, $\omega_n$ is the Matsubara frequency for bosons, $g$ is the interaction strength and $\alpha$ is the strength of the dissipation. This model obviously breaks the Galilean invariance because of the last term. 

\begin{figure}
    \centering
    \includegraphics[width=\columnwidth]{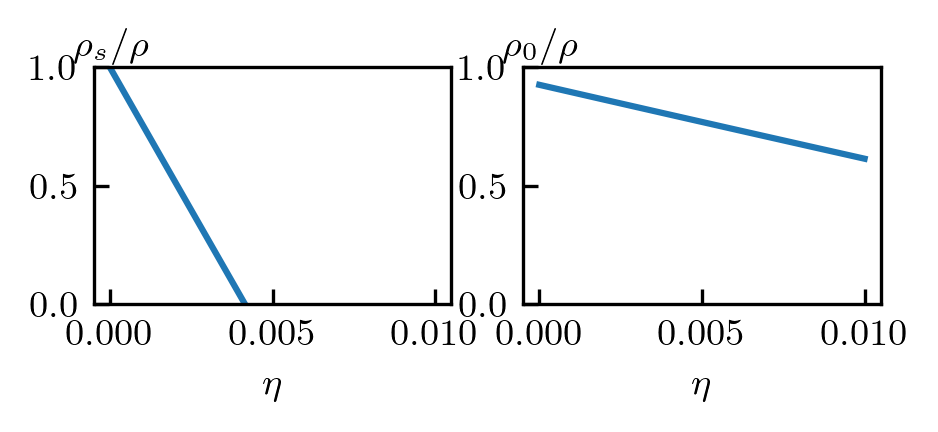}
    \caption{$\rho_s/\rho$ and $\rho_0/\rho$ obtained from Eqs. (\ref{sfdensity}) and (\ref{normalexp}). The parameters are $\epsilon_c/(\rho g)=900,\, \omega_c/(\rho g)=1000,\, m g = 1$.}
    \label{rhosrho0fig}
\end{figure}

We calculated the superfluid density by the Bogoliubov approximation, i.e., substitute $\psi=\sqrt{\rho_0}+\phi$ and $\bar{\psi}=\sqrt{\rho_0}+\bar{\phi}$ and retained the terms up to quadratic order in $\phi,\bar{\phi}$. From the general argument \cite{nozieres_theory_2018,griffin_excitations_1993}, the normal component $\rho_n = \rho - \rho_s$ can be obtained from the transverse current-current response function $\chi^t(\omega,\bm{q})$ as $\rho_n/m = \lim_{\bm{q}\to 0}\chi^t(0,\bm{q})$. In the imaginary time formalism, $\chi^t(\omega,\bm{q})$ can be calculated from the analytic continuation from the time ordered correlation function. As is noted in Ref. \onlinecite{keeling_superfluid_2011}, the contribution to the transverse current current correlation function at one-loop order is given by the bubble diagram and here the expression is the same as the one given in Ref. \onlinecite{keeling_superfluid_2011}:
\begin{align*}
    \frac{\rho_n}{m}=\lim_{\bm{q}\to 0}\int \frac{d\epsilon_{\bm{k}}}{2\pi}
    \frac{d\omega}{2\pi}
    \frac{ \epsilon_{\bm{k}}}{4 i}\text{tr}[\sigma_3 G^K_{\omega,\bm{k}}\sigma_3
    (G^R_{\omega,\bm{k}+\bm{q}}+G^A_{\omega,\bm{k}-\bm{q}})], \numberthis \label{sfdensity}
\end{align*}
where we assumed the two dimensional system. The green functions are given as,
\begin{align*}
 G^{R/A}_{\omega,\bm{k}}
&=\frac{1}{\omega^2-\omega_{\bm{k}}^2 \pm 2i\eta\omega\epsilon_{\bm{k}}}\\
&\times\begin{pmatrix}
 \omega+\epsilon_{\bm{k}}+g\rho_0 \mp i \eta\omega & -g\rho_0 \pm i\eta\omega\\
 -g\rho_0 \pm i\eta\omega &  -\omega+\epsilon_{\bm{k}}+g\rho_0 \mp i\eta\omega
\end{pmatrix},
\end{align*}
and $G^K_{\omega,\bm{k}} = \coth(\beta\omega/2) [G^R_{\omega,\bm{k}}-G^A_{\omega,\bm{k}}]$,
where $\eta = 2\rho_0\alpha$. The number density can be calculated from the $(1,1)$ component in the Nambu space of the lesser Green function:
\begin{align*}
\rho &= \rho_0 + \frac{1}{V}\sum_{\bm{k}}i (G^{<}_{\bm{k}}(t=0))_{11} \\
    &= \rho_0 + \frac{1}{V}\sum_{\bm{k}}\int_{-\infty}^{\infty} \frac{d\omega}{2\pi} n_B(\omega)i((G^R_{\omega,\bm{k}})_{11}-(G^A_{\omega,\bm{k}})_{11})
\numberthis \label{normalexp}
\end{align*}
From now on, we consider the zero temperature case, where the destruction of the superfluidity comes purely from the dissipation. The finite temperature case can be treated in a similar manner.
We introduce the cutoff for $\eta$ as $\eta\,\Theta(\omega_c^2-\omega^2)$, where $\Theta(x)$ is the step function. We also introduced the cutoff for the energy $\epsilon_{\bm{k}}$ at $\epsilon_c$, and choose $\epsilon_c<\omega_c$, so that the whole energy spectrum of the system is coupled to the heat bath. To calculate $\rho_s$, we regard $\eta$ as a control parameter, calculate $\rho_0$ as a function of $\eta$ and then calculate $\rho_s(\eta,\rho_0(\eta))$. The result of the calculation is shown in Fig. \ref{rhosrho0fig}. We can see that $\rho_s$ rapidly decreases and vanishes so the superfluidity is destroyed by the dissipation. This kind of behavior is also shown in Ref. \onlinecite{huang_hard-sphere_1992}, where the authors discussed the destruction of the superfluid by the static impurity potential, which is in contrast to our system where the translation symmetry is preserved but the time non-local action breaks the Galilean invariance.

From Fig. \ref{sfdensity}, we can see that, at the critical $\eta$ where $\rho_s=0$, $\rho_0$ remains finite. This behavior is similar to the system with disorder \cite{huang_hard-sphere_1992,lopatin_thermodynamics_2002}, but the depletion of $\rho_0$ is large in this parameter region, so our one-loop calculation cannot decide whether or not $\rho_0$ is finite at the critical point. In fact, assuming the smooth behavior of the single particle Green function at the critical point, the Josephson relation \cite{josephson_relation_1966,muller_josephson_2015} requires that both $\rho_0$ and $\rho_s$ becomes zero. In spite of this uncertainty, we believe that the transition to the phase with $\rho_s=0$ in this model remains intact, as is supported by our numerical calculation in a model with the different source of the Galilean symmetry breaking.

{\it Discussion.--- }
At a moderately clean regime $1/\tau,k_B T \gg \Delta^2/\epsilon_F$ \cite{kopnin_theory_2009}, where $\tau$ is the relaxation time, the particle-hole excitation at the normal core \cite{caroli_bound_1964,stone_magnus_1996} can be regarded as a heat bath with a continuum spectrum, so we can regard Eq. (\ref{dissaction}) as a model for the vortices with normal core in that regime.
Since the motion of vortex induces the resistivity \cite{bardeen_theory_1965} and the density of vortices is proportional to the magnetic field, we expect a giant magnetoresistance, as is observed experimentally \cite{saito_quantum_2018,kapitulnik_colloquium:_2019}.
For a weak magnetic field, the resistivity can be much smaller than 
the quantum resistance $h/(4 e^2)$.
We note that the long range interaction between the vortices does not spoil our scenario if we include the effect of the screening
\cite{feigelman_two-dimensional_1993,foldy_charged_1961,hore_dielectric_1975,lee_dielectric_1975}.

We also speculate that the effect of the normal core or dissipation discussed above affects the phase transition associated with the proliferation of the vortices, i.e., the transition not associated with the magnetic field. The point is that, if we extend the above dissipative action to the closed loop in the space-time, in the parameter region where the typical size of the vortex ring in the space-time is macroscopic, the exchange process between the rings is still suppressed from the same reason as above. Therefore, we expect a different phase compared to the usual proliferation of vortices in the bosonic superfluid.

In summary, we have shown both analytically and numerically that the presence of the heat bath, which is the continuous degrees of freedom, drastically affects the thermodynamic phase realized by the bosons. Our first model is in a first-quantized form, and we discussed the reduction of the superfluid density because of the modification of the {\it off-diagonal} density matrix of each particle in the presence of dissipation. We also numerically showed the reduction of the superfluid density at finite temperature to show the strong support for our scenario. Our second model is a second-quantized field theoretical model, and we calculated the superfluid density from the transverse current-current correlation function at zero temperature. 

\begin{acknowledgments}
We thank H. Ishizuka, A. Mishchenko and M. Ueda for fruitful discussions. This work was supported by JST CREST Grant (JPMJCR1874 and JPMJCR16F1) and JSPS KAKENHI (JP18H03676, JP26103006 and JP18J21329).
\end{acknowledgments}


\begin{thebibliography}{58}%
\makeatletter
\providecommand \@ifxundefined [1]{%
 \@ifx{#1\undefined}
}%
\providecommand \@ifnum [1]{%
 \ifnum #1\expandafter \@firstoftwo
 \else \expandafter \@secondoftwo
 \fi
}%
\providecommand \@ifx [1]{%
 \ifx #1\expandafter \@firstoftwo
 \else \expandafter \@secondoftwo
 \fi
}%
\providecommand \natexlab [1]{#1}%
\providecommand \enquote  [1]{``#1''}%
\providecommand \bibnamefont  [1]{#1}%
\providecommand \bibfnamefont [1]{#1}%
\providecommand \citenamefont [1]{#1}%
\providecommand \href@noop [0]{\@secondoftwo}%
\providecommand \href [0]{\begingroup \@sanitize@url \@href}%
\providecommand \@href[1]{\@@startlink{#1}\@@href}%
\providecommand \@@href[1]{\endgroup#1\@@endlink}%
\providecommand \@sanitize@url [0]{\catcode `\\12\catcode `\$12\catcode
  `\&12\catcode `\#12\catcode `\^12\catcode `\_12\catcode `\%12\relax}%
\providecommand \@@startlink[1]{}%
\providecommand \@@endlink[0]{}%
\providecommand \url  [0]{\begingroup\@sanitize@url \@url }%
\providecommand \@url [1]{\endgroup\@href {#1}{\urlprefix }}%
\providecommand \urlprefix  [0]{URL }%
\providecommand \Eprint [0]{\href }%
\providecommand \doibase [0]{http://dx.doi.org/}%
\providecommand \selectlanguage [0]{\@gobble}%
\providecommand \bibinfo  [0]{\@secondoftwo}%
\providecommand \bibfield  [0]{\@secondoftwo}%
\providecommand \translation [1]{[#1]}%
\providecommand \BibitemOpen [0]{}%
\providecommand \bibitemStop [0]{}%
\providecommand \bibitemNoStop [0]{.\EOS\space}%
\providecommand \EOS [0]{\spacefactor3000\relax}%
\providecommand \BibitemShut  [1]{\csname bibitem#1\endcsname}%
\let\auto@bib@innerbib\@empty
\bibitem [{\citenamefont {Kapitulnik}\ \emph {et~al.}(2019)\citenamefont
  {Kapitulnik}, \citenamefont {Kivelson},\ and\ \citenamefont
  {Spivak}}]{kapitulnik_colloquium:_2019}%
  \BibitemOpen
  \bibfield  {author} {\bibinfo {author} {\bibfnamefont {A.}~\bibnamefont
  {Kapitulnik}}, \bibinfo {author} {\bibfnamefont {S.~A.}\ \bibnamefont
  {Kivelson}}, \ and\ \bibinfo {author} {\bibfnamefont {B.}~\bibnamefont
  {Spivak}},\ }\href {\doibase 10.1103/RevModPhys.91.011002} {\bibfield
  {journal} {\bibinfo  {journal} {Reviews of Modern Physics}\ }\textbf
  {\bibinfo {volume} {91}},\ \bibinfo {pages} {011002} (\bibinfo {year}
  {2019})}\BibitemShut {NoStop}%
\bibitem [{\citenamefont {Dalidovich}\ and\ \citenamefont
  {Phillips}(2000)}]{dalidovich_fluctuation_2000}%
  \BibitemOpen
  \bibfield  {author} {\bibinfo {author} {\bibfnamefont {D.}~\bibnamefont
  {Dalidovich}}\ and\ \bibinfo {author} {\bibfnamefont {P.}~\bibnamefont
  {Phillips}},\ }\href {\doibase 10.1103/PhysRevLett.84.737} {\bibfield
  {journal} {\bibinfo  {journal} {Physical Review Letters}\ }\textbf {\bibinfo
  {volume} {84}},\ \bibinfo {pages} {737} (\bibinfo {year} {2000})}\BibitemShut
  {NoStop}%
\bibitem [{\citenamefont {Zhu}\ \emph {et~al.}(2015)\citenamefont {Zhu},
  \citenamefont {Chen},\ and\ \citenamefont {Varma}}]{zhu_local_2015}%
  \BibitemOpen
  \bibfield  {author} {\bibinfo {author} {\bibfnamefont {L.}~\bibnamefont
  {Zhu}}, \bibinfo {author} {\bibfnamefont {Y.}~\bibnamefont {Chen}}, \ and\
  \bibinfo {author} {\bibfnamefont {C.~M.}\ \bibnamefont {Varma}},\ }\href
  {\doibase 10.1103/PhysRevB.91.205129} {\bibfield  {journal} {\bibinfo
  {journal} {Physical Review B}\ }\textbf {\bibinfo {volume} {91}},\ \bibinfo
  {pages} {205129} (\bibinfo {year} {2015})}\BibitemShut {NoStop}%
\bibitem [{\citenamefont {Zhu}\ \emph {et~al.}(2016)\citenamefont {Zhu},
  \citenamefont {Hou},\ and\ \citenamefont {Varma}}]{zhu_quantum_2016}%
  \BibitemOpen
  \bibfield  {author} {\bibinfo {author} {\bibfnamefont {L.}~\bibnamefont
  {Zhu}}, \bibinfo {author} {\bibfnamefont {C.}~\bibnamefont {Hou}}, \ and\
  \bibinfo {author} {\bibfnamefont {C.~M.}\ \bibnamefont {Varma}},\ }\href
  {\doibase 10.1103/PhysRevB.94.235156} {\bibfield  {journal} {\bibinfo
  {journal} {Physical Review B}\ }\textbf {\bibinfo {volume} {94}},\ \bibinfo
  {pages} {235156} (\bibinfo {year} {2016})}\BibitemShut {NoStop}%
\bibitem [{\citenamefont {Hou}\ and\ \citenamefont
  {Varma}(2016)}]{hou_phase_2016}%
  \BibitemOpen
  \bibfield  {author} {\bibinfo {author} {\bibfnamefont {C.}~\bibnamefont
  {Hou}}\ and\ \bibinfo {author} {\bibfnamefont {C.~M.}\ \bibnamefont
  {Varma}},\ }\href {\doibase 10.1103/PhysRevB.94.201101} {\bibfield  {journal}
  {\bibinfo  {journal} {Physical Review B}\ }\textbf {\bibinfo {volume} {94}},\
  \bibinfo {pages} {201101(R)} (\bibinfo {year} {2016})}\BibitemShut {NoStop}%
\bibitem [{\citenamefont {Chakravarty}\ \emph {et~al.}(1986)\citenamefont
  {Chakravarty}, \citenamefont {Ingold}, \citenamefont {Kivelson},\ and\
  \citenamefont {Luther}}]{chakravarty_onset_1986}%
  \BibitemOpen
  \bibfield  {author} {\bibinfo {author} {\bibfnamefont {S.}~\bibnamefont
  {Chakravarty}}, \bibinfo {author} {\bibfnamefont {G.-L.}\ \bibnamefont
  {Ingold}}, \bibinfo {author} {\bibfnamefont {S.}~\bibnamefont {Kivelson}}, \
  and\ \bibinfo {author} {\bibfnamefont {A.}~\bibnamefont {Luther}},\ }\href
  {\doibase 10.1103/PhysRevLett.56.2303} {\bibfield  {journal} {\bibinfo
  {journal} {Physical Review Letters}\ }\textbf {\bibinfo {volume} {56}},\
  \bibinfo {pages} {2303} (\bibinfo {year} {1986})}\BibitemShut {NoStop}%
\bibitem [{\citenamefont {Chakravarty}\ \emph {et~al.}(1988)\citenamefont
  {Chakravarty}, \citenamefont {Ingold}, \citenamefont {Kivelson},\ and\
  \citenamefont {Zimanyi}}]{chakravarty_quantum_1988}%
  \BibitemOpen
  \bibfield  {author} {\bibinfo {author} {\bibfnamefont {S.}~\bibnamefont
  {Chakravarty}}, \bibinfo {author} {\bibfnamefont {G.-L.}\ \bibnamefont
  {Ingold}}, \bibinfo {author} {\bibfnamefont {S.}~\bibnamefont {Kivelson}}, \
  and\ \bibinfo {author} {\bibfnamefont {G.}~\bibnamefont {Zimanyi}},\ }\href
  {\doibase 10.1103/PhysRevB.37.3283} {\bibfield  {journal} {\bibinfo
  {journal} {Physical Review B}\ }\textbf {\bibinfo {volume} {37}},\ \bibinfo
  {pages} {3283} (\bibinfo {year} {1988})}\BibitemShut {NoStop}%
\bibitem [{\citenamefont {Fisher}\ and\ \citenamefont
  {Grinstein}(1988)}]{fisher_quantum_1988}%
  \BibitemOpen
  \bibfield  {author} {\bibinfo {author} {\bibfnamefont {M.~P.~A.}\
  \bibnamefont {Fisher}}\ and\ \bibinfo {author} {\bibfnamefont
  {G.}~\bibnamefont {Grinstein}},\ }\href {\doibase 10.1103/PhysRevLett.60.208}
  {\bibfield  {journal} {\bibinfo  {journal} {Physical Review Letters}\
  }\textbf {\bibinfo {volume} {60}},\ \bibinfo {pages} {208} (\bibinfo {year}
  {1988})}\BibitemShut {NoStop}%
\bibitem [{\citenamefont {Fisher}\ \emph {et~al.}(1989)\citenamefont {Fisher},
  \citenamefont {Weichman}, \citenamefont {Grinstein},\ and\ \citenamefont
  {Fisher}}]{fisher_boson_1989}%
  \BibitemOpen
  \bibfield  {author} {\bibinfo {author} {\bibfnamefont {M.~P.~A.}\
  \bibnamefont {Fisher}}, \bibinfo {author} {\bibfnamefont {P.~B.}\
  \bibnamefont {Weichman}}, \bibinfo {author} {\bibfnamefont {G.}~\bibnamefont
  {Grinstein}}, \ and\ \bibinfo {author} {\bibfnamefont {D.~S.}\ \bibnamefont
  {Fisher}},\ }\href {\doibase 10.1103/PhysRevB.40.546} {\bibfield  {journal}
  {\bibinfo  {journal} {Physical Review B}\ }\textbf {\bibinfo {volume} {40}},\
  \bibinfo {pages} {546} (\bibinfo {year} {1989})}\BibitemShut {NoStop}%
\bibitem [{\citenamefont {Sachdev}(2011)}]{sachdev_quantum_2011}%
  \BibitemOpen
  \bibfield  {author} {\bibinfo {author} {\bibfnamefont {S.}~\bibnamefont
  {Sachdev}},\ }\href@noop {} {{\selectlanguage {English}\emph {\bibinfo
  {title} {Quantum {Phase} {Transitions}}}}}\ (\bibinfo  {publisher} {Cambridge
  University Press},\ \bibinfo {address} {Cambridge},\ \bibinfo {year}
  {2011})\BibitemShut {NoStop}%
\bibitem [{\citenamefont {Cai}\ \emph {et~al.}(2014)\citenamefont {Cai},
  \citenamefont {Schollw\"ock},\ and\ \citenamefont
  {Pollet}}]{cai_identifying_2014}%
  \BibitemOpen
  \bibfield  {author} {\bibinfo {author} {\bibfnamefont {Z.}~\bibnamefont
  {Cai}}, \bibinfo {author} {\bibfnamefont {U.}~\bibnamefont
  {Schollw\"ock}}, \ and\ \bibinfo {author} {\bibfnamefont
  {L.}~\bibnamefont {Pollet}},\ }\href {\doibase
  10.1103/PhysRevLett.113.260403} {\bibfield  {journal} {\bibinfo  {journal}
  {Physical Review Letters}\ }\textbf {\bibinfo {volume} {113}},\ \bibinfo
  {pages} {260403} (\bibinfo {year} {2014})}\BibitemShut {NoStop}%
\bibitem [{\citenamefont {Greiter}\ \emph {et~al.}(1989)\citenamefont
  {Greiter}, \citenamefont {Wilczek},\ and\ \citenamefont
  {Witten}}]{greiter_hydrodynamic_1989}%
  \BibitemOpen
  \bibfield  {author} {\bibinfo {author} {\bibfnamefont {M.}~\bibnamefont
  {Greiter}}, \bibinfo {author} {\bibfnamefont {F.}~\bibnamefont {Wilczek}}, \
  and\ \bibinfo {author} {\bibfnamefont {E.}~\bibnamefont {Witten}},\ }\href
  {\doibase 10.1142/S0217984989001400} {\bibfield  {journal} {\bibinfo
  {journal} {Mod.Phys.Lett.}\ }\textbf {\bibinfo {volume} {B3}},\ \bibinfo
  {pages} {903} (\bibinfo {year} {1989})}\BibitemShut {NoStop}%
\bibitem [{\citenamefont {Son}\ and\ \citenamefont
  {Wingate}(2006)}]{son_general_2006}%
  \BibitemOpen
  \bibfield  {author} {\bibinfo {author} {\bibfnamefont {D.}~\bibnamefont
  {Son}}\ and\ \bibinfo {author} {\bibfnamefont {M.}~\bibnamefont {Wingate}},\
  }\href {\doibase 10.1016/j.aop.2005.11.001} {\bibfield  {journal} {\bibinfo
  {journal} {Annals of Physics}\ }\textbf {\bibinfo {volume} {321}},\ \bibinfo
  {pages} {197} (\bibinfo {year} {2006})}\BibitemShut {NoStop}%
\bibitem [{\citenamefont {Aitchison}\ \emph {et~al.}(1995)\citenamefont
  {Aitchison}, \citenamefont {Ao}, \citenamefont {Thouless},\ and\
  \citenamefont {Zhu}}]{aitchison_effective_1995}%
  \BibitemOpen
  \bibfield  {author} {\bibinfo {author} {\bibfnamefont {I.~J.~R.}\
  \bibnamefont {Aitchison}}, \bibinfo {author} {\bibfnamefont {P.}~\bibnamefont
  {Ao}}, \bibinfo {author} {\bibfnamefont {D.~J.}\ \bibnamefont {Thouless}}, \
  and\ \bibinfo {author} {\bibfnamefont {X.-M.}\ \bibnamefont {Zhu}},\ }\href
  {\doibase 10.1103/PhysRevB.51.6531} {\bibfield  {journal} {\bibinfo
  {journal} {Physical Review B}\ }\textbf {\bibinfo {volume} {51}},\ \bibinfo
  {pages} {6531} (\bibinfo {year} {1995})}\BibitemShut {NoStop}%
\bibitem [{\citenamefont {Aitchison}\ \emph {et~al.}(2000)\citenamefont
  {Aitchison}, \citenamefont {Metikas},\ and\ \citenamefont
  {Lee}}]{aitchison_finite-temperature_2000}%
  \BibitemOpen
  \bibfield  {author} {\bibinfo {author} {\bibfnamefont {I.~J.~R.}\
  \bibnamefont {Aitchison}}, \bibinfo {author} {\bibfnamefont {G.}~\bibnamefont
  {Metikas}}, \ and\ \bibinfo {author} {\bibfnamefont {D.~J.}\ \bibnamefont
  {Lee}},\ }\href {\doibase 10.1103/PhysRevB.62.6638} {\bibfield  {journal}
  {\bibinfo  {journal} {Physical Review B}\ }\textbf {\bibinfo {volume} {62}},\
  \bibinfo {pages} {6638} (\bibinfo {year} {2000})}\BibitemShut {NoStop}%
\bibitem [{\citenamefont {Popov}(1987)}]{popov_functional_1987}%
  \BibitemOpen
  \bibfield  {author} {\bibinfo {author} {\bibfnamefont {V.~N.}\ \bibnamefont
  {Popov}},\ }\href@noop {} {{\selectlanguage {English}\emph {\bibinfo {title}
  {Functional integrals and collective excitations}}}},\ Cambridge monographs
  on mathematical physics\ (\bibinfo  {publisher} {Cambridge Univ. Press},\
  \bibinfo {address} {Cambridge},\ \bibinfo {year} {1987})\BibitemShut
  {NoStop}%
\bibitem [{\citenamefont {Stoof}\ \emph {et~al.}(2009)\citenamefont {Stoof},
  \citenamefont {Dickerscheid},\ and\ \citenamefont
  {Gubbels}}]{stoof_ultracold_2009}%
  \BibitemOpen
  \bibfield  {author} {\bibinfo {author} {\bibfnamefont {H.~T.~C.}\
  \bibnamefont {Stoof}}, \bibinfo {author} {\bibfnamefont {D.~B.~M.}\
  \bibnamefont {Dickerscheid}}, \ and\ \bibinfo {author} {\bibfnamefont
  {K.}~\bibnamefont {Gubbels}},\ }\href@noop {} {{\selectlanguage
  {English}\emph {\bibinfo {title} {Ultracold quantum fields}}}}\ (\bibinfo
  {publisher} {Springer},\ \bibinfo {address} {Berlin},\ \bibinfo {year}
  {2009})\BibitemShut {NoStop}%
\bibitem [{\citenamefont {Imada}\ \emph {et~al.}(1998)\citenamefont {Imada},
  \citenamefont {Fujimori},\ and\ \citenamefont
  {Tokura}}]{imada_metal-insulator_1998}%
  \BibitemOpen
  \bibfield  {author} {\bibinfo {author} {\bibfnamefont {M.}~\bibnamefont
  {Imada}}, \bibinfo {author} {\bibfnamefont {A.}~\bibnamefont {Fujimori}}, \
  and\ \bibinfo {author} {\bibfnamefont {Y.}~\bibnamefont {Tokura}},\ }\href
  {\doibase 10.1103/RevModPhys.70.1039} {\bibfield  {journal} {\bibinfo
  {journal} {Reviews of Modern Physics}\ }\textbf {\bibinfo {volume} {70}},\
  \bibinfo {pages} {1039} (\bibinfo {year} {1998})}\BibitemShut {NoStop}%
\bibitem [{\citenamefont {Huang}\ and\ \citenamefont
  {Meng}(1992)}]{huang_hard-sphere_1992}%
  \BibitemOpen
  \bibfield  {author} {\bibinfo {author} {\bibfnamefont {K.}~\bibnamefont
  {Huang}}\ and\ \bibinfo {author} {\bibfnamefont {H.-F.}\ \bibnamefont
  {Meng}},\ }\href {\doibase 10.1103/PhysRevLett.69.644} {\bibfield  {journal}
  {\bibinfo  {journal} {Physical Review Letters}\ }\textbf {\bibinfo {volume}
  {69}},\ \bibinfo {pages} {644} (\bibinfo {year} {1992})}\BibitemShut
  {NoStop}%
\bibitem [{\citenamefont {Lopatin}\ and\ \citenamefont
  {Vinokur}(2002)}]{lopatin_thermodynamics_2002}%
  \BibitemOpen
  \bibfield  {author} {\bibinfo {author} {\bibfnamefont {A.~V.}\ \bibnamefont
  {Lopatin}}\ and\ \bibinfo {author} {\bibfnamefont {V.~M.}\ \bibnamefont
  {Vinokur}},\ }\href {\doibase 10.1103/PhysRevLett.88.235503} {\bibfield
  {journal} {\bibinfo  {journal} {Physical Review Letters}\ }\textbf {\bibinfo
  {volume} {88}},\ \bibinfo {pages} {235503} (\bibinfo {year}
  {2002})}\BibitemShut {NoStop}%
\bibitem [{\citenamefont {Feigelman}\ \emph {et~al.}(1993)\citenamefont
  {Feigelman}, \citenamefont {Geshkenbein}, \citenamefont {Ioffe},\ and\
  \citenamefont {Larkin}}]{feigelman_two-dimensional_1993}%
  \BibitemOpen
  \bibfield  {author} {\bibinfo {author} {\bibfnamefont {M.~V.}\ \bibnamefont
  {Feigelman}}, \bibinfo {author} {\bibfnamefont {V.~B.}\ \bibnamefont
  {Geshkenbein}}, \bibinfo {author} {\bibfnamefont {L.~B.}\ \bibnamefont
  {Ioffe}}, \ and\ \bibinfo {author} {\bibfnamefont {A.~I.}\ \bibnamefont
  {Larkin}},\ }\href {\doibase 10.1103/PhysRevB.48.16641} {\bibfield  {journal}
  {\bibinfo  {journal} {Physical Review B}\ }\textbf {\bibinfo {volume} {48}},\
  \bibinfo {pages} {16641} (\bibinfo {year} {1993})}\BibitemShut {NoStop}%
\bibitem [{\citenamefont {Weiss}(2012)}]{weiss_quantum_2012}%
  \BibitemOpen
  \bibfield  {author} {\bibinfo {author} {\bibfnamefont {U.}~\bibnamefont
  {Weiss}},\ }\href {\doibase 10.1142/8334} {\emph {\bibinfo {title} {Quantum
  {Dissipative} {Systems}}}},\ \bibinfo {edition} {4th}\ ed.\ (\bibinfo
  {publisher} {WORLD SCIENTIFIC},\ \bibinfo {year} {2012})\BibitemShut
  {NoStop}%
\bibitem [{\citenamefont {Feynman}(1953{\natexlab{a}})}]{feynman_atomic_1953}%
  \BibitemOpen
  \bibfield  {author} {\bibinfo {author} {\bibfnamefont {R.~P.}\ \bibnamefont
  {Feynman}},\ }\href {\doibase 10.1103/PhysRev.91.1291} {\bibfield  {journal}
  {\bibinfo  {journal} {Physical Review}\ }\textbf {\bibinfo {volume} {91}},\
  \bibinfo {pages} {1291} (\bibinfo {year} {1953}{\natexlab{a}})}\BibitemShut
  {NoStop}%
\bibitem [{\citenamefont {Feynman}(1954)}]{feynman_atomic_1954}%
  \BibitemOpen
  \bibfield  {author} {\bibinfo {author} {\bibfnamefont {R.~P.}\ \bibnamefont
  {Feynman}},\ }\href {\doibase 10.1103/PhysRev.94.262} {\bibfield  {journal}
  {\bibinfo  {journal} {Physical Review}\ }\textbf {\bibinfo {volume} {94}},\
  \bibinfo {pages} {262} (\bibinfo {year} {1954})}\BibitemShut {NoStop}%
\bibitem [{\citenamefont {Kopnin}(2009)}]{kopnin_theory_2009}%
  \BibitemOpen
  \bibfield  {author} {\bibinfo {author} {\bibfnamefont {N.~B.}\ \bibnamefont
  {Kopnin}},\ }\href@noop {} {{\selectlanguage {English}\emph {\bibinfo {title}
  {Theory of nonequilibrium superconductivity}}}},\ \bibinfo {series}
  {International series of monographs on physics}\ No.\ \bibinfo {number}
  {110}\ (\bibinfo  {publisher} {Oxford Univ. Press},\ \bibinfo {address}
  {Oxford},\ \bibinfo {year} {2009})\BibitemShut {NoStop}%
\bibitem [{\citenamefont {Caldeira}\ and\ \citenamefont
  {Leggett}(1983)}]{caldeira_quantum_1983}%
  \BibitemOpen
  \bibfield  {author} {\bibinfo {author} {\bibfnamefont {A.~O.}\ \bibnamefont
  {Caldeira}}\ and\ \bibinfo {author} {\bibfnamefont {A.~J.}\ \bibnamefont
  {Leggett}},\ }\href {\doibase 10.1016/0003-4916(83)90202-6} {\bibfield
  {journal} {\bibinfo  {journal} {Annals of Physics}\ }\textbf {\bibinfo
  {volume} {149}},\ \bibinfo {pages} {374} (\bibinfo {year}
  {1983})}\BibitemShut {NoStop}%
\bibitem [{\citenamefont {Nagaosa}(1999)}]{nagaosa_quantum_1999}%
  \BibitemOpen
  \bibfield  {author} {\bibinfo {author} {\bibfnamefont {N.}~\bibnamefont
  {Nagaosa}},\ }\href {https://www.springer.com/jp/book/9783540655374}
  {{\selectlanguage {English}\emph {\bibinfo {title} {Quantum {Field} {Theory}
  in {Condensed} {Matter} {Physics}}}}},\ Theoretical and {Mathematical}
  {Physics}\ (\bibinfo  {publisher} {Springer-Verlag},\ \bibinfo {address}
  {Berlin Heidelberg},\ \bibinfo {year} {1999})\BibitemShut {NoStop}%
\bibitem [{\citenamefont {Duarte}\ and\ \citenamefont
  {Caldeira}(2006)}]{duarte_effective_2006}%
  \BibitemOpen
  \bibfield  {author} {\bibinfo {author} {\bibfnamefont {O.~S.}\ \bibnamefont
  {Duarte}}\ and\ \bibinfo {author} {\bibfnamefont {A.~O.}\ \bibnamefont
  {Caldeira}},\ }\href {\doibase 10.1103/PhysRevLett.97.250601} {\bibfield
  {journal} {\bibinfo  {journal} {Physical Review Letters}\ }\textbf {\bibinfo
  {volume} {97}},\ \bibinfo {pages} {250601} (\bibinfo {year}
  {2006})}\BibitemShut {NoStop}%
\bibitem [{\citenamefont
  {Feynman}(1953{\natexlab{b}})}]{feynman_$ensuremathlambda$-transition_1953}%
  \BibitemOpen
  \bibfield  {author} {\bibinfo {author} {\bibfnamefont {R.~P.}\ \bibnamefont
  {Feynman}},\ }\href {\doibase 10.1103/PhysRev.90.1116.2} {\bibfield
  {journal} {\bibinfo  {journal} {Physical Review}\ }\textbf {\bibinfo {volume}
  {90}},\ \bibinfo {pages} {1116} (\bibinfo {year}
  {1953}{\natexlab{b}})}\BibitemShut {NoStop}%
\bibitem [{\citenamefont {Pollock}\ and\ \citenamefont
  {Ceperley}(1987)}]{pollock_path-integral_1987}%
  \BibitemOpen
  \bibfield  {author} {\bibinfo {author} {\bibfnamefont {E.~L.}\ \bibnamefont
  {Pollock}}\ and\ \bibinfo {author} {\bibfnamefont {D.~M.}\ \bibnamefont
  {Ceperley}},\ }\href {\doibase 10.1103/PhysRevB.36.8343} {\bibfield
  {journal} {\bibinfo  {journal} {Physical Review B}\ }\textbf {\bibinfo
  {volume} {36}},\ \bibinfo {pages} {8343} (\bibinfo {year}
  {1987})}\BibitemShut {NoStop}%
\bibitem [{\citenamefont {Ceperley}(1995)}]{ceperley_path_1995}%
  \BibitemOpen
  \bibfield  {author} {\bibinfo {author} {\bibfnamefont {D.~M.}\ \bibnamefont
  {Ceperley}},\ }\href {\doibase 10.1103/RevModPhys.67.279} {\bibfield
  {journal} {\bibinfo  {journal} {Reviews of Modern Physics}\ }\textbf
  {\bibinfo {volume} {67}},\ \bibinfo {pages} {279} (\bibinfo {year}
  {1995})}\BibitemShut {NoStop}%
\bibitem [{\citenamefont {Apenko}(1999)}]{apenko_critical_1999}%
  \BibitemOpen
  \bibfield  {author} {\bibinfo {author} {\bibfnamefont {S.~M.}\ \bibnamefont
  {Apenko}},\ }\href {\doibase 10.1103/PhysRevB.60.3052} {\bibfield  {journal}
  {\bibinfo  {journal} {Physical Review B}\ }\textbf {\bibinfo {volume} {60}},\
  \bibinfo {pages} {3052} (\bibinfo {year} {1999})}\BibitemShut {NoStop}%
\bibitem [{Note1()}]{Note1}%
  \BibitemOpen
  \bibinfo {note} {We note that there is a similar problem in the treatment of
  the effective mass of polaron \cite
  {feynman_slow_1955,khandekar_polaron_1988}.}\BibitemShut {Stop}%
\bibitem [{\citenamefont {Boninsegni}\ \emph
  {et~al.}(2006{\natexlab{a}})\citenamefont {Boninsegni}, \citenamefont
  {Prokof'ev},\ and\ \citenamefont {Svistunov}}]{boninsegni_worm_2006-1}%
  \BibitemOpen
  \bibfield  {author} {\bibinfo {author} {\bibfnamefont {M.}~\bibnamefont
  {Boninsegni}}, \bibinfo {author} {\bibfnamefont {N.}~\bibnamefont
  {Prokof'ev}}, \ and\ \bibinfo {author} {\bibfnamefont {B.}~\bibnamefont
  {Svistunov}},\ }\href {\doibase 10.1103/PhysRevLett.96.070601} {\bibfield
  {journal} {\bibinfo  {journal} {Physical Review Letters}\ }\textbf {\bibinfo
  {volume} {96}},\ \bibinfo {pages} {070601} (\bibinfo {year}
  {2006}{\natexlab{a}})}\BibitemShut {NoStop}%
\bibitem [{\citenamefont {Boninsegni}\ \emph
  {et~al.}(2006{\natexlab{b}})\citenamefont {Boninsegni}, \citenamefont
  {Prokof'ev},\ and\ \citenamefont {Svistunov}}]{boninsegni_worm_2006}%
  \BibitemOpen
  \bibfield  {author} {\bibinfo {author} {\bibfnamefont {M.}~\bibnamefont
  {Boninsegni}}, \bibinfo {author} {\bibfnamefont {N.~V.}\ \bibnamefont
  {Prokof'ev}}, \ and\ \bibinfo {author} {\bibfnamefont {B.~V.}\ \bibnamefont
  {Svistunov}},\ }\href {\doibase 10.1103/PhysRevE.74.036701} {\bibfield
  {journal} {\bibinfo  {journal} {Physical Review E}\ }\textbf {\bibinfo
  {volume} {74}},\ \bibinfo {pages} {036701} (\bibinfo {year}
  {2006}{\natexlab{b}})}\BibitemShut {NoStop}%
\bibitem [{\citenamefont {Mezzacapo}\ and\ \citenamefont
  {Boninsegni}(2006)}]{mezzacapo_superfluidity_2006}%
  \BibitemOpen
  \bibfield  {author} {\bibinfo {author} {\bibfnamefont {F.}~\bibnamefont
  {Mezzacapo}}\ and\ \bibinfo {author} {\bibfnamefont {M.}~\bibnamefont
  {Boninsegni}},\ }\href {\doibase 10.1103/PhysRevLett.97.045301} {\bibfield
  {journal} {\bibinfo  {journal} {Physical Review Letters}\ }\textbf {\bibinfo
  {volume} {97}},\ \bibinfo {pages} {045301} (\bibinfo {year}
  {2006})}\BibitemShut {NoStop}%
\bibitem [{\citenamefont {Mezzacapo}\ and\ \citenamefont
  {Boninsegni}(2007)}]{mezzacapo_structure_2007}%
  \BibitemOpen
  \bibfield  {author} {\bibinfo {author} {\bibfnamefont {F.}~\bibnamefont
  {Mezzacapo}}\ and\ \bibinfo {author} {\bibfnamefont {M.}~\bibnamefont
  {Boninsegni}},\ }\href {\doibase 10.1103/PhysRevA.75.033201} {\bibfield
  {journal} {\bibinfo  {journal} {Physical Review A}\ }\textbf {\bibinfo
  {volume} {75}},\ \bibinfo {pages} {033201} (\bibinfo {year}
  {2007})}\BibitemShut {NoStop}%
\bibitem [{\citenamefont {Aziz}\ \emph {et~al.}(1979)\citenamefont {Aziz},
  \citenamefont {Nain}, \citenamefont {Carley}, \citenamefont {Taylor},\ and\
  \citenamefont {McConville}}]{aziz_accurate_1979}%
  \BibitemOpen
  \bibfield  {author} {\bibinfo {author} {\bibfnamefont {R.~A.}\ \bibnamefont
  {Aziz}}, \bibinfo {author} {\bibfnamefont {V.~P.~S.}\ \bibnamefont {Nain}},
  \bibinfo {author} {\bibfnamefont {J.~S.}\ \bibnamefont {Carley}}, \bibinfo
  {author} {\bibfnamefont {W.~L.}\ \bibnamefont {Taylor}}, \ and\ \bibinfo
  {author} {\bibfnamefont {G.~T.}\ \bibnamefont {McConville}},\ }\href
  {\doibase 10.1063/1.438007} {\bibfield  {journal} {\bibinfo  {journal} {The
  Journal of Chemical Physics}\ }\textbf {\bibinfo {volume} {70}},\ \bibinfo
  {pages} {4330} (\bibinfo {year} {1979})}\BibitemShut {NoStop}%
\bibitem [{\citenamefont {Ambegaokar}\ and\ \citenamefont
  {Troyer}(2010)}]{ambegaokar_estimating_2010}%
  \BibitemOpen
  \bibfield  {author} {\bibinfo {author} {\bibfnamefont {V.}~\bibnamefont
  {Ambegaokar}}\ and\ \bibinfo {author} {\bibfnamefont {M.}~\bibnamefont
  {Troyer}},\ }\href {\doibase 10.1119/1.3247985} {\bibfield  {journal}
  {\bibinfo  {journal} {American Journal of Physics}\ }\textbf {\bibinfo
  {volume} {78}},\ \bibinfo {pages} {150} (\bibinfo {year} {2010})}\BibitemShut
  {NoStop}%
\bibitem [{\citenamefont {Chin}(1997)}]{chin_symplectic_1997}%
  \BibitemOpen
  \bibfield  {author} {\bibinfo {author} {\bibfnamefont {S.~A.}\ \bibnamefont
  {Chin}},\ }\href {\doibase 10.1016/S0375-9601(97)00003-0} {\bibfield
  {journal} {\bibinfo  {journal} {Physics Letters A}\ }\textbf {\bibinfo
  {volume} {226}},\ \bibinfo {pages} {344} (\bibinfo {year}
  {1997})}\BibitemShut {NoStop}%
\bibitem [{\citenamefont {Werner}\ \emph
  {et~al.}(2005{\natexlab{a}})\citenamefont {Werner}, \citenamefont {Troyer},\
  and\ \citenamefont {Sachdev}}]{werner_quantum_2005}%
  \BibitemOpen
  \bibfield  {author} {\bibinfo {author} {\bibfnamefont {P.}~\bibnamefont
  {Werner}}, \bibinfo {author} {\bibfnamefont {M.}~\bibnamefont {Troyer}}, \
  and\ \bibinfo {author} {\bibfnamefont {S.}~\bibnamefont {Sachdev}},\ }\href
  {\doibase 10.1143/JPSJS.74S.67} {\bibfield  {journal} {\bibinfo  {journal}
  {Journal of the Physical Society of Japan}\ }\textbf {\bibinfo {volume}
  {74}},\ \bibinfo {pages} {67} (\bibinfo {year}
  {2005}{\natexlab{a}})}\BibitemShut {NoStop}%
\bibitem [{\citenamefont {Werner}\ \emph
  {et~al.}(2005{\natexlab{b}})\citenamefont {Werner}, \citenamefont
  {V\"olker}, \citenamefont {Troyer},\ and\ \citenamefont
  {Chakravarty}}]{werner_phase_2005}%
  \BibitemOpen
  \bibfield  {author} {\bibinfo {author} {\bibfnamefont {P.}~\bibnamefont
  {Werner}}, \bibinfo {author} {\bibfnamefont {K.}~\bibnamefont
  {V\"olker}}, \bibinfo {author} {\bibfnamefont {M.}~\bibnamefont
  {Troyer}}, \ and\ \bibinfo {author} {\bibfnamefont {S.}~\bibnamefont
  {Chakravarty}},\ }\href {\doibase 10.1103/PhysRevLett.94.047201} {\bibfield
  {journal} {\bibinfo  {journal} {Physical Review Letters}\ }\textbf {\bibinfo
  {volume} {94}},\ \bibinfo {pages} {047201} (\bibinfo {year}
  {2005}{\natexlab{b}})}\BibitemShut {NoStop}%
\bibitem [{Note2()}]{Note2}%
  \BibitemOpen
  \bibinfo {note} {See supplemental material for the details of the off
  diagonal density matrix and the numerical calculation in two
  dimension.}\BibitemShut {Stop}%
\bibitem [{\citenamefont {Glyde}(2017)}]{glyde_excitations_2017}%
  \BibitemOpen
  \bibfield  {author} {\bibinfo {author} {\bibfnamefont {H.~R.}\ \bibnamefont
  {Glyde}},\ }\href {\doibase 10.1088/1361-6633/aa7f90} {\bibfield  {journal}
  {\bibinfo  {journal} {Reports on Progress in Physics}\ }\textbf {\bibinfo
  {volume} {81}},\ \bibinfo {pages} {014501} (\bibinfo {year}
  {2017})}\BibitemShut {NoStop}%
\bibitem [{\citenamefont {Nozieres}(2018)}]{nozieres_theory_2018}%
  \BibitemOpen
  \bibfield  {author} {\bibinfo {author} {\bibfnamefont {P.}~\bibnamefont
  {Nozieres}},\ }\href {\doibase 10.1201/9780429495717} {{\selectlanguage
  {English}\emph {\bibinfo {title} {Theory {Of} {Quantum} {Liquids}}}}}\
  (\bibinfo  {publisher} {CRC Press},\ \bibinfo {year} {2018})\BibitemShut
  {NoStop}%
\bibitem [{\citenamefont {Griffin}(1993)}]{griffin_excitations_1993}%
  \BibitemOpen
  \bibfield  {author} {\bibinfo {author} {\bibfnamefont {A.}~\bibnamefont
  {Griffin}},\ }\href {\doibase 10.1017/CBO9780511524257} {\emph {\bibinfo
  {title} {Excitations in a {Bose}-condensed liquid}}}\ (\bibinfo  {publisher}
  {Cambridge University Press},\ \bibinfo {address} {Cambridge},\ \bibinfo
  {year} {1993})\BibitemShut {NoStop}%
\bibitem [{\citenamefont {Keeling}(2011)}]{keeling_superfluid_2011}%
  \BibitemOpen
  \bibfield  {author} {\bibinfo {author} {\bibfnamefont {J.}~\bibnamefont
  {Keeling}},\ }\href {\doibase 10.1103/PhysRevLett.107.080402} {\bibfield
  {journal} {\bibinfo  {journal} {Physical Review Letters}\ }\textbf {\bibinfo
  {volume} {107}},\ \bibinfo {pages} {080402} (\bibinfo {year}
  {2011})}\BibitemShut {NoStop}%
\bibitem [{\citenamefont {Josephson}(1966)}]{josephson_relation_1966}%
  \BibitemOpen
  \bibfield  {author} {\bibinfo {author} {\bibfnamefont {B.~D.}\ \bibnamefont
  {Josephson}},\ }\href {\doibase 10.1016/0031-9163(66)90088-6} {\bibfield
  {journal} {\bibinfo  {journal} {Physics Letters}\ }\textbf {\bibinfo {volume}
  {21}},\ \bibinfo {pages} {608} (\bibinfo {year} {1966})}\BibitemShut
  {NoStop}%
\bibitem [{\citenamefont {M^^c3^^bcller}(2015)}]{muller_josephson_2015}%
  \BibitemOpen
  \bibfield  {author} {\bibinfo {author} {\bibfnamefont {C.~A.}\ \bibnamefont
  {M^^c3^^bcller}},\ }\href {\doibase 10.1103/PhysRevA.91.023602} {\bibfield
  {journal} {\bibinfo  {journal} {Physical Review A}\ }\textbf {\bibinfo
  {volume} {91}},\ \bibinfo {pages} {023602} (\bibinfo {year}
  {2015})}\BibitemShut {NoStop}%
\bibitem [{\citenamefont {Caroli}\ \emph {et~al.}(1964)\citenamefont {Caroli},
  \citenamefont {De~Gennes},\ and\ \citenamefont
  {Matricon}}]{caroli_bound_1964}%
  \BibitemOpen
  \bibfield  {author} {\bibinfo {author} {\bibfnamefont {C.}~\bibnamefont
  {Caroli}}, \bibinfo {author} {\bibfnamefont {P.~G.}\ \bibnamefont
  {De~Gennes}}, \ and\ \bibinfo {author} {\bibfnamefont {J.}~\bibnamefont
  {Matricon}},\ }\href {\doibase 10.1016/0031-9163(64)90375-0} {\bibfield
  {journal} {\bibinfo  {journal} {Physics Letters}\ }\textbf {\bibinfo {volume}
  {9}},\ \bibinfo {pages} {307} (\bibinfo {year} {1964})}\BibitemShut {NoStop}%
\bibitem [{\citenamefont {Stone}(1996)}]{stone_magnus_1996}%
  \BibitemOpen
  \bibfield  {author} {\bibinfo {author} {\bibfnamefont {M.}~\bibnamefont
  {Stone}},\ }\href {\doibase 10.1103/PhysRevB.53.16573} {\bibfield  {journal}
  {\bibinfo  {journal} {Physical Review B}\ }\textbf {\bibinfo {volume} {53}},\
  \bibinfo {pages} {16573} (\bibinfo {year} {1996})}\BibitemShut {NoStop}%
\bibitem [{\citenamefont {Bardeen}\ and\ \citenamefont
  {Stephen}(1965)}]{bardeen_theory_1965}%
  \BibitemOpen
  \bibfield  {author} {\bibinfo {author} {\bibfnamefont {J.}~\bibnamefont
  {Bardeen}}\ and\ \bibinfo {author} {\bibfnamefont {M.~J.}\ \bibnamefont
  {Stephen}},\ }\href {\doibase 10.1103/PhysRev.140.A1197} {\bibfield
  {journal} {\bibinfo  {journal} {Physical Review}\ }\textbf {\bibinfo {volume}
  {140}},\ \bibinfo {pages} {A1197} (\bibinfo {year} {1965})}\BibitemShut
  {NoStop}%
\bibitem [{\citenamefont {Saito}\ \emph {et~al.}(2018)\citenamefont {Saito},
  \citenamefont {Nojima},\ and\ \citenamefont {Iwasa}}]{saito_quantum_2018}%
  \BibitemOpen
  \bibfield  {author} {\bibinfo {author} {\bibfnamefont {Y.}~\bibnamefont
  {Saito}}, \bibinfo {author} {\bibfnamefont {T.}~\bibnamefont {Nojima}}, \
  and\ \bibinfo {author} {\bibfnamefont {Y.}~\bibnamefont {Iwasa}},\ }\href
  {\doibase 10.1038/s41467-018-03275-z} {\bibfield  {journal} {\bibinfo
  {journal} {Nature Communications}\ }\textbf {\bibinfo {volume} {9}},\
  \bibinfo {pages} {778} (\bibinfo {year} {2018})}\BibitemShut {NoStop}%
\bibitem [{\citenamefont {Foldy}(1961)}]{foldy_charged_1961}%
  \BibitemOpen
  \bibfield  {author} {\bibinfo {author} {\bibfnamefont {L.~L.}\ \bibnamefont
  {Foldy}},\ }\href {\doibase 10.1103/PhysRev.124.649} {\bibfield  {journal}
  {\bibinfo  {journal} {Physical Review}\ }\textbf {\bibinfo {volume} {124}},\
  \bibinfo {pages} {649} (\bibinfo {year} {1961})}\BibitemShut {NoStop}%
\bibitem [{\citenamefont {Hore}\ and\ \citenamefont
  {Frankel}(1975)}]{hore_dielectric_1975}%
  \BibitemOpen
  \bibfield  {author} {\bibinfo {author} {\bibfnamefont {S.~R.}\ \bibnamefont
  {Hore}}\ and\ \bibinfo {author} {\bibfnamefont {N.~E.}\ \bibnamefont
  {Frankel}},\ }\href {\doibase 10.1103/PhysRevB.12.2619} {\bibfield  {journal}
  {\bibinfo  {journal} {Physical Review B}\ }\textbf {\bibinfo {volume} {12}},\
  \bibinfo {pages} {2619} (\bibinfo {year} {1975})}\BibitemShut {NoStop}%
\bibitem [{\citenamefont {Lee}(1975)}]{lee_dielectric_1975}%
  \BibitemOpen
  \bibfield  {author} {\bibinfo {author} {\bibfnamefont {J.~C.}\ \bibnamefont
  {Lee}},\ }\href {\doibase 10.1103/PhysRevB.12.2644} {\bibfield  {journal}
  {\bibinfo  {journal} {Physical Review B}\ }\textbf {\bibinfo {volume} {12}},\
  \bibinfo {pages} {2644} (\bibinfo {year} {1975})}\BibitemShut {NoStop}%
\bibitem [{\citenamefont {Feynman}(1955)}]{feynman_slow_1955}%
  \BibitemOpen
  \bibfield  {author} {\bibinfo {author} {\bibfnamefont {R.~P.}\ \bibnamefont
  {Feynman}},\ }\href {\doibase 10.1103/PhysRev.97.660} {\bibfield  {journal}
  {\bibinfo  {journal} {Physical Review}\ }\textbf {\bibinfo {volume} {97}},\
  \bibinfo {pages} {660} (\bibinfo {year} {1955})}\BibitemShut {NoStop}%
\bibitem [{\citenamefont {Khandekar}\ \emph {et~al.}(1988)\citenamefont
  {Khandekar}, \citenamefont {Bhagwat},\ and\ \citenamefont
  {Lawande}}]{khandekar_polaron_1988}%
  \BibitemOpen
  \bibfield  {author} {\bibinfo {author} {\bibfnamefont {D.~C.}\ \bibnamefont
  {Khandekar}}, \bibinfo {author} {\bibfnamefont {K.~V.}\ \bibnamefont
  {Bhagwat}}, \ and\ \bibinfo {author} {\bibfnamefont {S.~V.}\ \bibnamefont
  {Lawande}},\ }\href {\doibase 10.1103/PhysRevB.37.3085} {\bibfield  {journal}
  {\bibinfo  {journal} {Physical Review B}\ }\textbf {\bibinfo {volume} {37}},\
  \bibinfo {pages} {3085} (\bibinfo {year} {1988})}\BibitemShut {NoStop}%
\end{thebibliography}

\begin{thebibliography}{1}%
\makeatletter
\providecommand \@ifxundefined [1]{%
 \@ifx{#1\undefined}
}%
\providecommand \@ifnum [1]{%
 \ifnum #1\expandafter \@firstoftwo
 \else \expandafter \@secondoftwo
 \fi
}%
\providecommand \@ifx [1]{%
 \ifx #1\expandafter \@firstoftwo
 \else \expandafter \@secondoftwo
 \fi
}%
\providecommand \natexlab [1]{#1}%
\providecommand \enquote  [1]{``#1''}%
\providecommand \bibnamefont  [1]{#1}%
\providecommand \bibfnamefont [1]{#1}%
\providecommand \citenamefont [1]{#1}%
\providecommand \href@noop [0]{\@secondoftwo}%
\providecommand \href [0]{\begingroup \@sanitize@url \@href}%
\providecommand \@href[1]{\@@startlink{#1}\@@href}%
\providecommand \@@href[1]{\endgroup#1\@@endlink}%
\providecommand \@sanitize@url [0]{\catcode `\\12\catcode `\$12\catcode
  `\&12\catcode `\#12\catcode `\^12\catcode `\_12\catcode `\%12\relax}%
\providecommand \@@startlink[1]{}%
\providecommand \@@endlink[0]{}%
\providecommand \url  [0]{\begingroup\@sanitize@url \@url }%
\providecommand \@url [1]{\endgroup\@href {#1}{\urlprefix }}%
\providecommand \urlprefix  [0]{URL }%
\providecommand \Eprint [0]{\href }%
\providecommand \doibase [0]{http://dx.doi.org/}%
\providecommand \selectlanguage [0]{\@gobble}%
\providecommand \bibinfo  [0]{\@secondoftwo}%
\providecommand \bibfield  [0]{\@secondoftwo}%
\providecommand \translation [1]{[#1]}%
\providecommand \BibitemOpen [0]{}%
\providecommand \bibitemStop [0]{}%
\providecommand \bibitemNoStop [0]{.\EOS\space}%
\providecommand \EOS [0]{\spacefactor3000\relax}%
\providecommand \BibitemShut  [1]{\csname bibitem#1\endcsname}%
\let\auto@bib@innerbib\@empty
\bibitem [{\citenamefont {Weiss}(2012)}]{weiss_quantum_2012_supp}%
  \BibitemOpen
  \bibfield  {author} {\bibinfo {author} {\bibfnamefont {U.}~\bibnamefont
  {Weiss}},\ }\href {\doibase 10.1142/8334} {\emph {\bibinfo {title} {Quantum
  {Dissipative} {Systems}}}},\ \bibinfo {edition} {4th}\ ed.\ (\bibinfo
  {publisher} {WORLD SCIENTIFIC},\ \bibinfo {year} {2012})\BibitemShut
  {NoStop}%
\end{thebibliography}
%

\appendix
\setcounter{equation}{0}
\setcounter{figure}{0}
\setcounter{table}{0}
\makeatletter
\renewcommand{\theequation}{S\arabic{equation}}
\renewcommand{\thefigure}{S\arabic{figure}}
\renewcommand{\bibnumfmt}[1]{[S#1]}
\renewcommand{\citenumfont}[1]{S#1}
\begin{widetext}
\pagebreak

\begin{center}
\large{\textbf{Supplemental material for ``Suppression of superfluidity by dissipation \\
--- An application to failed superconductor''}}
\end{center}

\section{Off diagonal density matrix in three dimension}
Here, we will discuss the result of the numerical calculation of the off diagonal density matrix. The action used in the simulation is the same as the one in the main text:
\begin{align}
    S&=\int_0^{\beta} d\tau \left(\sum_i \frac{m}{2}\dot{\vec{r}}+\sum_{i>j}V_{i,j}\right)
    +\frac{\eta}{4\pi}
    \sum_i \int_0^{\beta}d\tau\int_0^{\beta}d\tau' \frac{\pi^2}{\beta^2}
    \left(\frac{\vec{r}_i(\tau)-\vec{r}_i(\tau')}{\sin\frac{\pi}{\beta}(\tau-\tau')}\right)^2. \label{dissaction_supp}
\end{align}
Here we will show the result for the single particle and the many particle ($N=64$) system in Fig. \ref{odmat}.

As we can see, as for the single particle case, the effect of dissipation appears in the decrease of the width of the Gaussian distribution. 
To see this, we showed the off diagonal density matrix for the single particle case obtained both from the numerical calculation and the analytical expression \cite{weiss_quantum_2012_supp}, 
\begin{align}
    \braket{p^2}&=\frac{M}{\beta}+ 2M \frac{\mu_1 \mu_2}{\mu_1-\mu_2} \left[\psi\left(1+\mu_1\beta\right)-\psi\left(1+\mu_2\beta\right)\right],\quad
    n(r) = \exp \left(-\frac{\braket{p^2}r^2}{2 \hbar^2}\right) \label{psquare}
\end{align}
where $\psi(x)$ is the digamma function, $\mu_{1/2}=\hbar(\omega_D \pm \sqrt{\omega_D^2-4\gamma\omega_D})/(4\pi)$. The cutoff for this expression is the Drude type cutoff, i.e., $J(\omega) = \eta \omega/(1+(\omega/\omega_D)^2)$, where $J(\omega)$ is the spectral function of the bath. Although the form of the cutoff in the numerical calculation is different from the Drude cutoff, we can see that the numerical and analytical result agrees well.

For the many particle case, the dissipation does not change the width very much, but it leads to the decrease of the condensate fraction $n_0$ as is discussed in the main text. We fitted the off diagonal density matrix with $\tilde{n}_0+(1-\tilde{n}_0)f(r)$, where $f(x)=\exp(-\bar{\alpha}_2 x^2/2!+\bar{\alpha}_4 x^4/4! - \bar{\alpha}_6 x^6/6!)$. The result is shown in Fig. \ref{odmat2}. Although the $\alpha_2$, which represents the second cumulant of the distribution, does not change drastically as a function of $\eta$, $\alpha_4$ and $\alpha_6$ decreases, which indicates that the distribution becomes more and more Gaussian-like distribution.
\begin{figure}[p]
    \centering
    \includegraphics{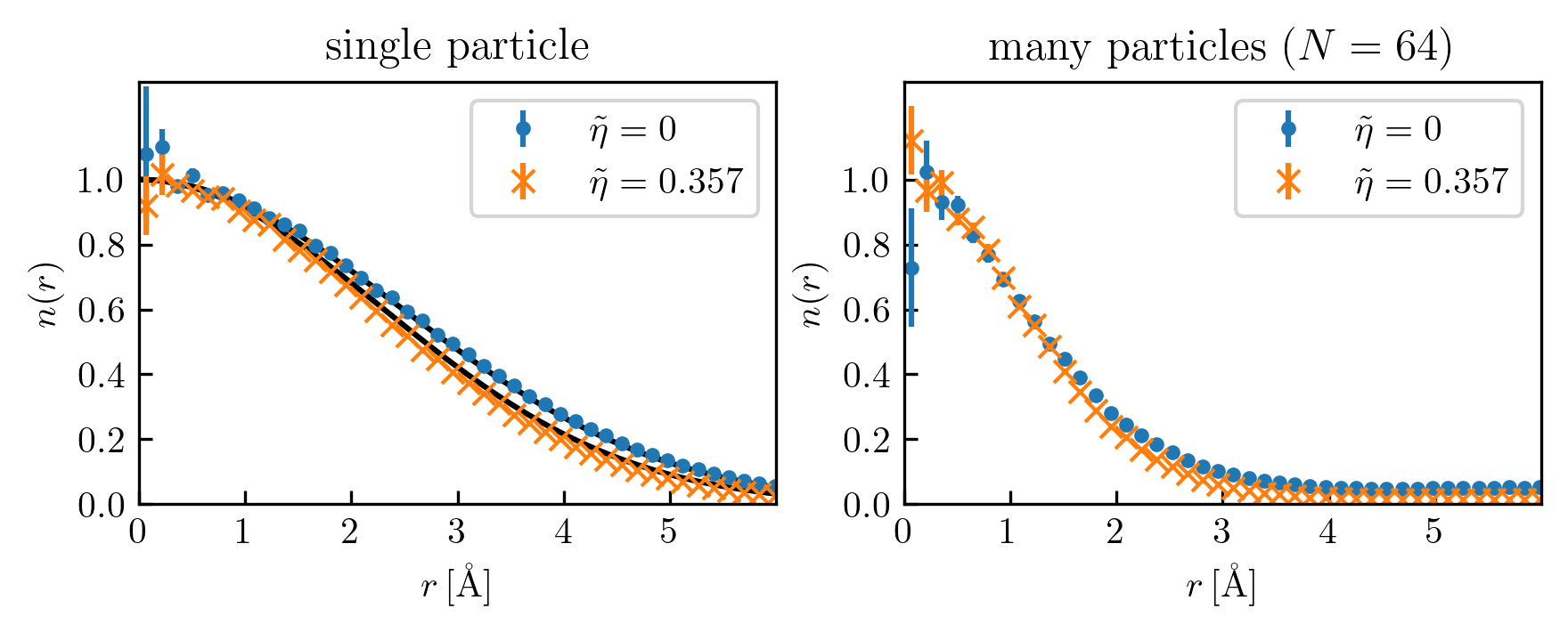}
    \caption{The off diagonal density matrix in three dimension for the single particle and the particles with interaction. The parameters are the same as the one in the main text: The number of particles is $64$, the density is at the saturated vapor pressure, the temperature is $T = 2 \,{\rm K}$, the imaginary time step is $5\times 10^{-3}\,{\rm K^{-1}}$, and the cutoff of the bath is set to be $\tau_c=0.2$. $\tilde{\eta}=\eta d^2$, where $d = 3.570 \, {\rm \AA}$. The solid lines for the single particle case is an analytical result based on the Eq. (\ref{psquare}).}
    \label{odmat}
\end{figure}
\begin{figure}
    \centering
    \includegraphics{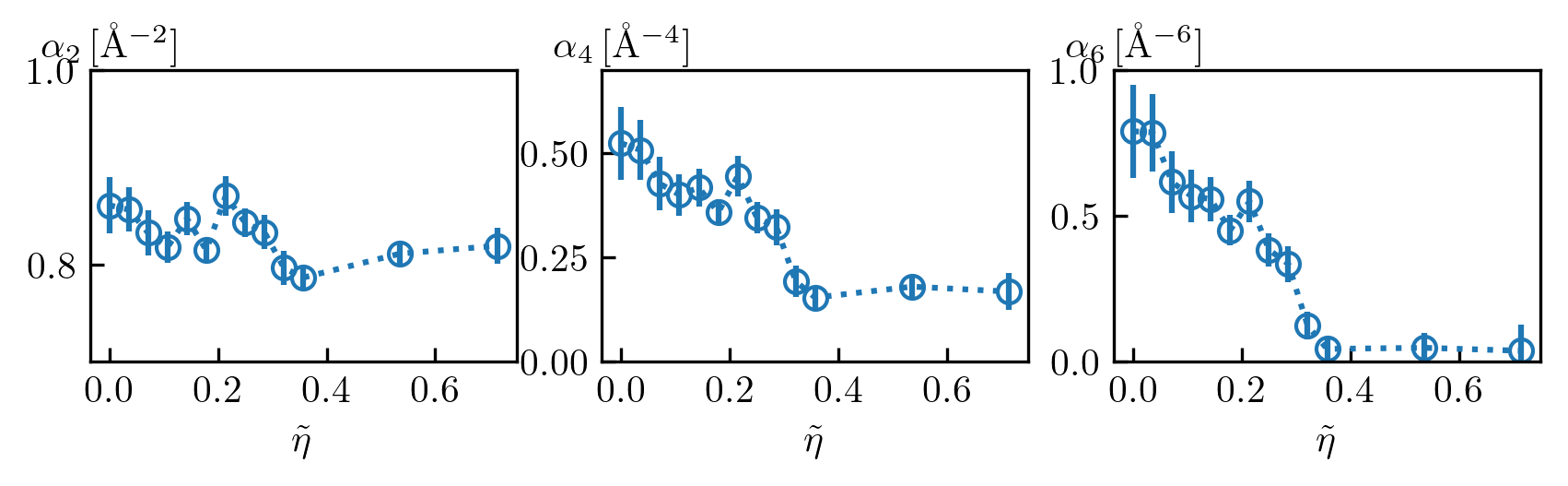}
    \caption{The parameters for the off diagonal density matrix in three dimension. The fitting function is $\tilde{n}_0+(1-\tilde{n}_0)f(r)$, where $f(x)=\exp(-\bar{\alpha}_2 x^2/2!+\bar{\alpha}_4 x^4/4! - \bar{\alpha}_6 x^6/6!)$.}
    \label{odmat2}
\end{figure}

\section{Numerical calculation in two dimension}
Here, we will show the suppression of the superfluidity in two dimension to show that our scenario for the suppression of superfluidity does not depend on the dimensionality of the system. We performed the quantum Monte Carlo in two dimension with the following parameters: the temperature $T = 0.5\,{\rm K}$; the number of particles $N=25$; the particle density $0.0432\,{\rm \AA^{-2}}$; the cutoff of the bath $\tau_c=0.05$; the imaginary time step $5\times 10^{-3}\,{\rm K^{-1}}$. The result is shown in Fig. \ref{odmat2d}. As we can see, the superfluid fraction decreases as a function of $\eta$. Also, the tail of the off diagonal density matrix is drastically suppressed in the presence of the dissipation.
\begin{figure}
    \centering
    \includegraphics{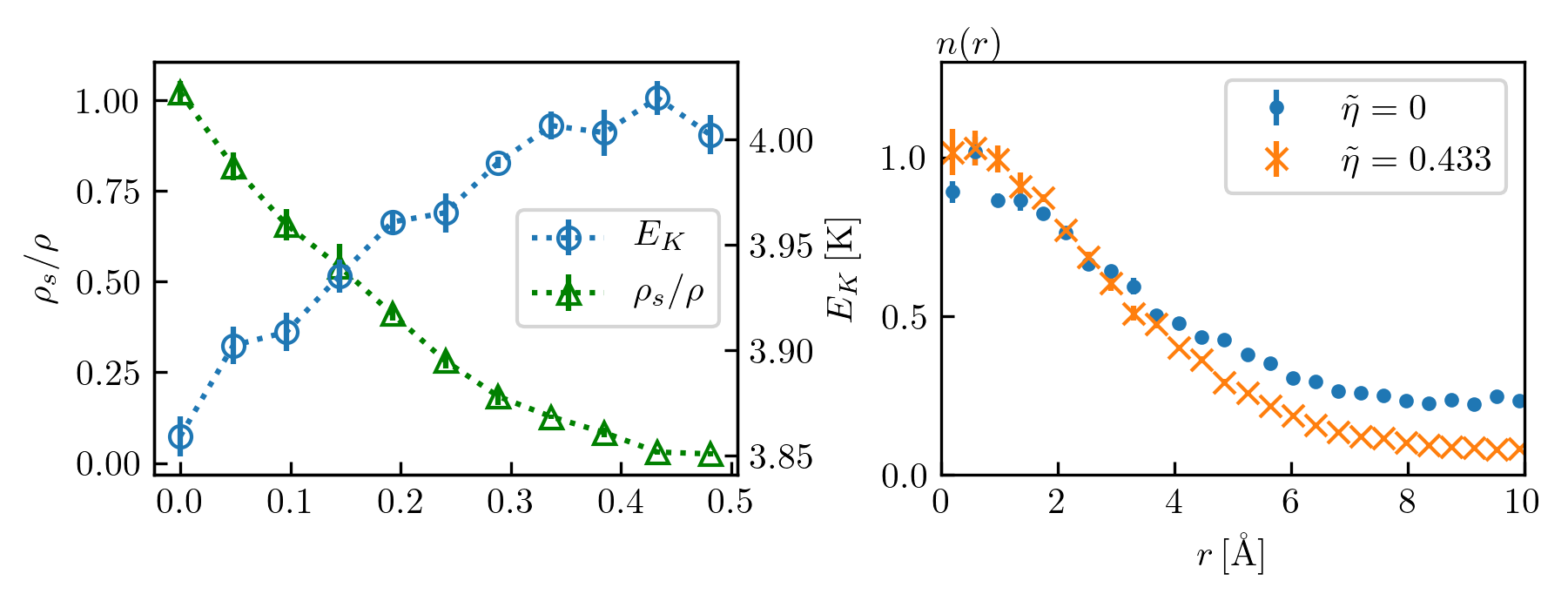}
    \caption{The kinetic energy $E_K$, the superfluid fraction $\rho_s$ and the off diagonal density matrix for the two dimensional system. $\tilde{\eta}=\eta d^2$, where $d = 4.811 \, {\rm \AA}$.}
    \label{odmat2d}
\end{figure}

\end{widetext}

\end{document}